\documentclass[conference]{IEEEtran}
\IEEEoverridecommandlockouts

\usepackage{cite}
\usepackage{amsmath,amssymb,amsfonts,amsthm}
\usepackage{textcomp}
\usepackage{graphicx}
\usepackage{xcolor}
\usepackage{hyperref}

\usepackage{algorithm}
\usepackage[noend]{algorithmic}

\usepackage[capitalize]{cleveref}
\usepackage{enumitem}

\usepackage{subfig}

\DeclareMathOperator*{\argmax}{arg\,max}

\newtheorem{definition}{Definition}
\newtheorem{example}{Example}
\newtheorem{lemma}{Lemma}
\newtheorem{theorem}{Theorem}

\begin{document}

\title{Opinion-aware Influence Maximization in Online Social Networks}

\author{\IEEEauthorblockN{Ying Wang$^\sharp$, Yanhao Wang$^\dagger$}
\IEEEauthorblockA{\textit{School of Data Science and Engineering, East China Normal University, Shanghai, China}\\
$^\sharp$\texttt{yingwang007@stu.ecnu.edu.cn} \quad $^\dagger$\texttt{yhwang@dase.ecnu.edu.cn}}}

\maketitle 

\begin{abstract}
  Influence maximization (IM) aims to find seed users on an online social network to maximize the spread of information about a target product through word-of-mouth propagation among all users.
  Prior IM methods mostly focus on maximizing the overall influence spread, which assumes that all users are potential customers of the product and that more exposure leads to higher benefits.
  However, in real-world scenarios, some users who dislike the product may express and spread negative opinions, damaging the product's reputation and lowering its profit.
  This paper investigates the opinion-aware influence maximization (OIM) problem, which finds a set of seed users to maximize the positive opinions toward the product while minimizing the negative opinions.
  We propose a novel algorithm for the OIM problem.
  Specifically, after obtaining the users with positive and negative opinions towards the product from historical data, we design a reverse reachable set-based method for opinion-aware influence estimation and a sandwich approximation algorithm for seed set selection.
  Despite the NP-hardness and non-submodularity of OIM, our algorithm achieves a data-dependent approximation factor for OIM.
  Experimental results on three real-world datasets demonstrate that our algorithm improves the spread of positive opinions while reducing the spread of negative opinions compared to existing methods.
\end{abstract}


\section{Introduction}
\label{sec:intro}

With the boom in social networking, the problem of influence maximization~\cite{DBLP:conf/kdd/KempeKT03} (IM) has been extensively studied over the last two decades for its wide applications in viral marketing~\cite{DBLP:conf/kdd/DomingosR01} and recommendation~\cite{DBLP:conf/www/Fan0LHZTY19}.
The vanilla IM problem aims to find a set of seed users from an online social network (OSN) to maximize the spread of information about a target product.
For example, when a company wants to launch a promotion campaign on an OSN, instead of promoting its product to all users at once, a more cost-effective strategy is to select a small set of influential users on the network as seeds to initiate the spread of promotion information, hoping that a larger number of users can be reached through word-of-mouth propagation along social connections.
However, the vanilla IM problem only considers maximizing the total exposure of the product to all users but ignores users' opinions.
In real-life scenarios, some users may express and spread negative opinions about the product, damaging its reputation and lowering its profit.
As an illustrative example, a singer may want to promote her/his new song to a group of audiences who are more likely to comment on it positively and spread it to their friends who may also like it while avoiding dispersing it to the ones who may express negative opinions about it.

Motivated by the above application, opinion-aware influence maximization~\cite{DBLP:conf/sdm/ChenCCKLRSWWY11, DBLP:conf/icdcs/ZhangDT13, DBLP:conf/sigmod/GalhotraAR16, DBLP:conf/kdd/LiuKY18, DBLP:conf/asunam/LuoLK19} (OIM) problems consider users' opinions towards specific items (e.g., products and events), which might be positive, neutral, or negative, and their goals are mostly to maximize the positive opinions while minimizing the negative opinions among all influenced users.
Although OIM has been extensively studied in the literature, prior work still has several limitations.
Most existing methods~\cite{DBLP:conf/sdm/ChenCCKLRSWWY11, DBLP:conf/icdcs/ZhangDT13, DBLP:conf/sigmod/GalhotraAR16} are item-unaware in the sense that they do not consider incorporating historical information into OIM.
They typically assume that user opinions are arbitrarily generated or formed merely from their neighbors' opinions based on some given opinion models.
Although a few studies~\cite{DBLP:conf/kdd/LiuKY18, DBLP:conf/asunam/LuoLK19} on OIM are based on historical data to estimate users' opinions towards the target product, their methods for opinion estimation and seed set selection are heuristics without performance guarantees.

\begin{figure}[t]
  \centering
  \includegraphics[width=0.99\linewidth]{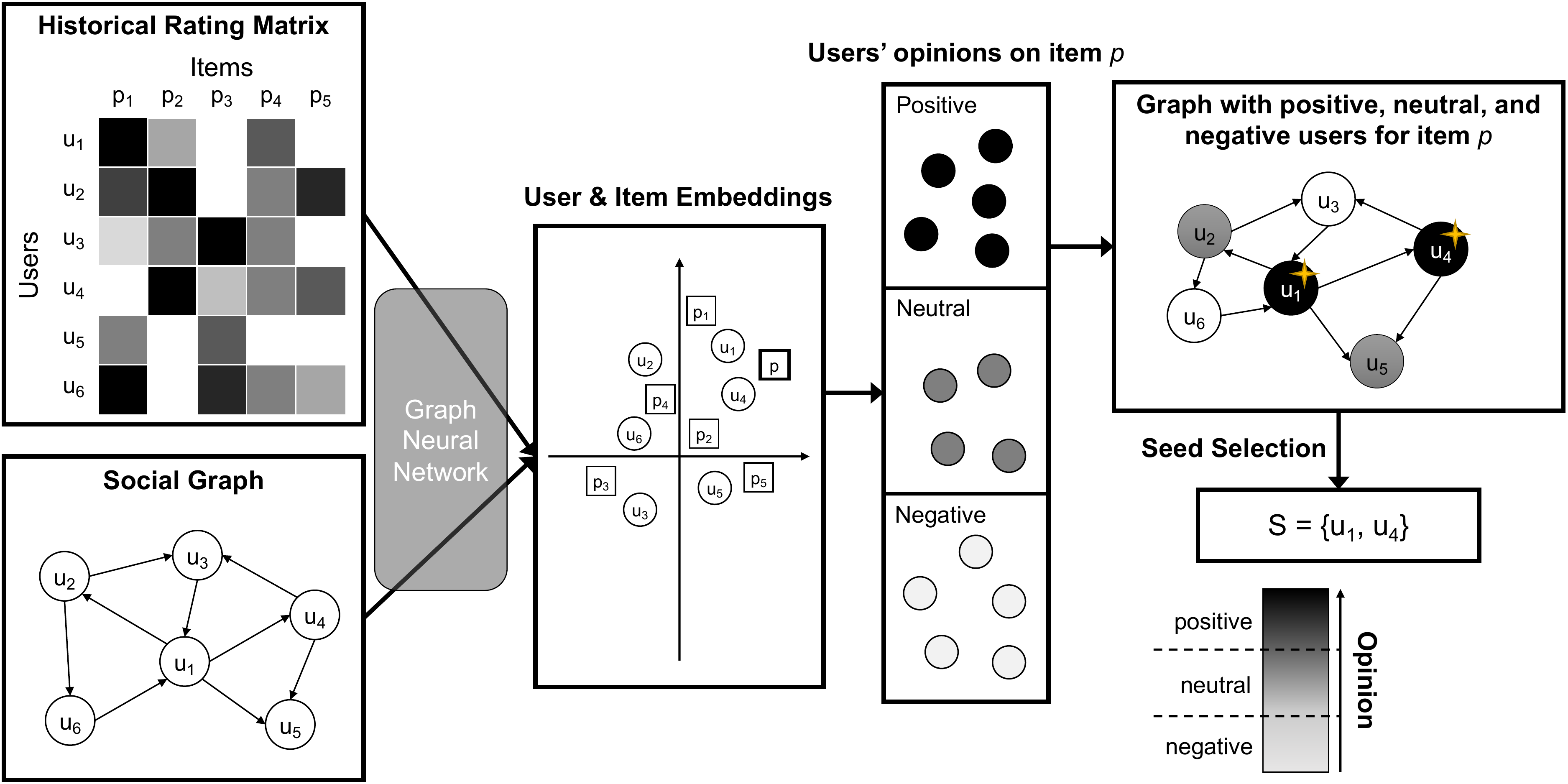}
  \caption{Illustration for opinion-aware IM in an OSN.}
  \label{fig:framework}
\end{figure}

In this work, we investigate a new opinion-aware influence maximization (OIM) problem on OSNs as illustrated in \cref{fig:framework}.
Specifically, we adopt any graph neural network model for recommendation~\cite{DBLP:conf/www/Fan0LHZTY19} that captures both users' historical rating information and social connections to learn latent vector representations of items (products) and users, from which the sets of users with positive, neutral, and negative opinions towards the target item are obtained.
Then, we generalize the independent cascade model to be opinion-aware to describe the information diffusion process of the item on the social graph.
Based on the above concepts, we formalize our OIM problem, aimed to find a set of $k$ seed users from the graph to maximize the difference between the influence spread on the positive and negative sets of users for the given item under the opinion-aware IC model.
We show that OIM is NP-hard and its objective function is non-monotone, neither submodular nor supermodular.
Consequently, the classic greedy algorithm for submodular maximization~\cite{DBLP:journals/mp/NemhauserWF78}, which has been widely used in existing IM methods~\cite{DBLP:conf/kdd/KempeKT03}, cannot provide any theoretical guarantee on OIM.
To address this issue, we propose a novel algorithm for OIM, which consists of an extended reverse reachable (RR) set-based method~\cite{DBLP:conf/soda/BorgsBCL14, DBLP:conf/sigmod/TangSX15} for opinion-aware influence estimation and a sandwich approximation strategy for greedy seed selection based on the upper- and lower-bound submodular functions of the objective function of OIM derived from the difference of submodular (DS) decomposition~\cite{DBLP:conf/uai/IyerB12}.
Our proposed algorithm runs in polynomial time and achieves a data-dependent approximation for OIM.
Finally, we conduct extensive experiments on three real-world datasets to compare our proposed algorithm with existing IM methods.
The results demonstrate that our algorithm significantly improves the spread of positive opinions about a target item and reduces the negative ones over existing IM methods while running in a comparable time.
Our main contributions are listed as follows:
\begin{itemize}[leftmargin=*]
  \item We formally define the opinion-aware influence maximization problem and show its theoretical hardness. (\cref{sec:def})
  \item We propose a data-dependent approximation algorithm consisting of RR set-based influence estimation and sandwich approximation-based seed selection for OIM. (\cref{sec:alg})
  \item We perform extensive experiments to verify the effectiveness and efficiency of our proposed algorithm. (\cref{sec:exp})
\end{itemize}

\section{Related Work}
\label{sec:literature}

In this section, we briefly discuss the literature related to this work, namely \emph{influence maximization} and \emph{opinion-aware influence maximization}.

\smallskip\noindent\textbf{Influence Maximization:}
There has been a vast amount of literature on influence maximization in OSNs (see~\cite{DBLP:journals/tkde/LiFWT18} for an extensive survey).
The IM problem was first formulated by Kempe \emph{et al.}~\cite{DBLP:conf/kdd/KempeKT03}, where two propagation models, i.e., the independent cascade (IC) model and the linear threshold (LT) model, were used to simulate the information diffusion process on social graphs.
They then showed the NP-hardness and submodularity of IM under IC and LT models.
Moreover, Chen \emph{et al.}~\cite{DBLP:conf/kdd/ChenWW10} proved that computing the influence spread under IC and LT models is \#P-hard.
Despite the theoretical hardness, a bunch of heuristic and approximation algorithms~\cite{DBLP:conf/kdd/ChenWY09, DBLP:conf/kdd/ChenWW10, DBLP:conf/soda/BorgsBCL14,DBLP:conf/aaai/OhsakaAYK14, DBLP:conf/sigmod/TangXS14, DBLP:conf/sigmod/TangSX15, DBLP:conf/sigmod/NguyenTD16, DBLP:conf/sigmod/Guo0WC20} were proposed to scale IM on massive graphs with millions of nodes.
However, all the above methods for vanilla IM problems do not consider using customized seeding schemes for different items.

\smallskip\noindent\textbf{Opinion-aware Influence Maximization:}
Unlike vanilla IM, opinion-aware IM (OIM) problems consider maximizing total influence spread while reducing negative opinions.
Chen \emph{et al.}~\cite{DBLP:conf/sdm/ChenCCKLRSWWY11} proposed an IC-N model by considering that negative opinions emerge randomly during the IC diffusion process as well as efficient IM methods under the IC-N model.
The OIM problems in~\cite{DBLP:conf/icdcs/ZhangDT13, DBLP:conf/sigmod/GalhotraAR16, DBLP:conf/kdd/LiuKY18, DBLP:conf/asunam/LuoLK19} were the most similar to ours since their goals were to maximize the total opinion in the form of the difference between positive and negative opinions under the IC or LT model.
Nonetheless, the OIM problems in~\cite{DBLP:conf/icdcs/ZhangDT13, DBLP:conf/sigmod/GalhotraAR16} still do not consider how to obtain user opinions from historical data.
In addition, most of them~\cite{DBLP:conf/icdcs/ZhangDT13, DBLP:conf/kdd/LiuKY18, DBLP:conf/asunam/LuoLK19} are specific for the LT model but cannot work with the more popular IC model.
Finally, their proposed OIM algorithms were all heuristics without performance guarantees.
Therefore, none of the above methods can work directly for the problem in this paper.

\section{Problem Formulation}
\label{sec:def}

Let $[n]$ denote a set of integers $ \{1, \ldots, n\} $.
An online social network (OSN) is represented as a directed graph $ G = \left( V, E, w \right) $, where $ V = \{ u_1, \ldots, u_n \} $ is a set of $n$ users (nodes), $ E \subseteq V \times V $ is a set of social links (edges) among users and $ e = (u, v) \in E $ is the edge from $u$ to $v$, and $ w: E \rightarrow \mathbb{R}^{+} $ assigns a nonnegative weight $w(e)$ to each edge $e \in E$.
We treat an undirected graph as a special case of a directed graph with symmetric directed edges.
We have a set of $m$ items (products) $ P = \{ p_1, \ldots, p_m \} $ associated with graph $ G $.
Let $ R = \{ r_{p u} \in \mathbb{R} \,|\, p \in P, u \in V \} $ be the set of observed ratings, where $ r_{p u} $ is the rating of user $ u \in V $ on item $ p \in P $.
All rating values are normalized to be nonnegative, and higher rating values indicate more positive opinions.
We assume that a graph neural network (GNN) based model for recommendation~\cite{DBLP:conf/www/Fan0LHZTY19} is used to learn latent vector representations $\mathbf{h}_u$ and $\mathbf{h}_p$ of each user $u \in V$ and item $p \in P$ from $R$ and $G$ to jointly capture the effects of both a user's interest and her/his neighborhoods' preferences on her/his rating on an item.
Then, the preference of user $u$ on item $p$ is estimated from their latent vector representations, i.e., $\hat{r}_{p u} = s(\mathbf{h}_{p}, \mathbf{h}_{u}) \in \mathbb{R}^{+}$, where the scoring function $s(\cdot, \cdot)$ can be the inner product, cosine, multi-layer perception, etc.
We define the opinion $o_{p u}$ of user $u$ on item $p$ to be ``positive'' ($1$), ``neutral'' ($0$), or ``negative'' ($-1$) based on the difference between the predicted rating value $\hat{r}_{pu}$ and the neutral rating value $r_0$ as follows:
\begin{equation}\label{eq:opinion}
  o_{p u} =
  \begin{cases}
    1, \quad \hat{r}_{p u} - r_0 \geq \tau \\
    0, \quad |\hat{r}_{p u} - r_0| < \tau \\
    -1, \quad r_0 - \hat{r}_{p u} \geq \tau
  \end{cases}
\end{equation}
where $\tau > 0$ decides the width of neutral zones.
In practice, we set the value of $r_0$ to the average or median of all (observed) rating values.
For any item $p$, the set $V$ of all users can be divided into three subsets based on their opinions on $p$, i.e., \emph{positive set} $ V^{+}_{p} = \{ u \in V \,|\, o_{p u} = 1 \} $, \emph{neutral set} $V^{0}_{p} = \{ u \in V \,|\, o_{p u} = 0 \} $, and \emph{negative set} $ V^{-}_{p} = \{ u \in V \,|\, o_{p u} = -1 \} $.

We use an opinion-aware variant of the independent cascade (OIC) model to describe the information propagation process on $G$, as the IC model and its variants have been widely used in the existing literature on influence and opinion maximization problems~\cite{DBLP:conf/kdd/KempeKT03, DBLP:conf/sdm/ChenCCKLRSWWY11}.
The influence diffusion process under the OIC model is described as follows.
There are four states, namely ``positively active'', ``neutrally active'', ``negatively active'', and ``inactive'', for each user to represent whether the user is affected by the promotion campaign (i.e., ``active'' vs.~``inactive'') and her/his opinion to the promoted item (i.e., ``positive'', ``neutral'', or ``negative'').
Initially, all the nodes in $ V $ are inactive.
Then, a set $ S \subseteq V $ of seed nodes are selected as initiators to promote item $p$ to others.
For each $ u \in S $, the state of $u$ will be set to either ``positively active'', or ``neutrally active'', or ``negatively active'' if $o_{pu} = 1$, $0$, or $-1$, respectively.
At step $t+1$, each node $u$ activated at step $t$ will try to activate each of its neighbors $v$ that is inactive at step $t$ along edge $e = (u, v)$ with probability $w(e)$.
If the activation is successful, $v$ will be (positively, neutral, or negatively) activated based on the value of $o_{pv}$.
Note that each node has only one chance to activate its neighbors.
The nodes activated at previous steps always remain active.
This process will terminate until no more nodes can be activated.
The total numbers of positive and negative nodes activated after the diffusion process on $G$ are defined as the \emph{positive influence spread} $\mathbb{E}[I^+_{G}(S)] = \sum_{u \in V^{+}_{p}} \Pr[S \leadsto u]$ and \emph{negative influence spread} $\mathbb{E}[I^-_{G}(S)] = \sum_{u \in V^{-}_{p}} \Pr[S \leadsto u]$ of item $p$, respectively, where $ \Pr[S \leadsto u] $ is the probability that $S$ activates $u$ under the OIC model on $G$.
We aim to maximize the positive influence while minimizing the negative influence for a given item.
We thus define the objective function $ \mathbb{E}[I_{G}(\cdot)] $ as the difference between the positive and negative influence spread functions, i.e.,
\begin{equation}\label{eq:obj}
  \mathbb{E}[I_{G}(S)] = \mathbb{E}[I^+_{G}(S)] - \mathbb{E}[I^-_{G}(S)].
\end{equation}
In this paper, we are concerned about the problem of selecting a set of \emph{seed nodes} to maximize $\mathbb{E}[I_{G}(\cdot)]$ in \cref{eq:obj}, which is formally defined by the \emph{opinion-aware influence maximization} (OIM) problem as follows.
\begin{definition}\label{def:iom}[Opinion-aware Influence Maximization (OIM)]
  For a graph $G$, a set $P$ of items, a set $R$ of ratings, a target item $p$, the OIC model on $G$, and the seed set size $k \in \mathbb{Z}^+$, select a set $S^*$ of $k$ seed nodes from $V$ such that $\mathbb{E}[I_{G}(S^*)]$ is maximized, i.e., $S^* = \argmax_{S \subseteq V, |S| = k} \mathbb{E}[I_{G}(S)]$.
\end{definition}

\begin{figure}[t]
  \centering
  \includegraphics[width=0.5\linewidth]{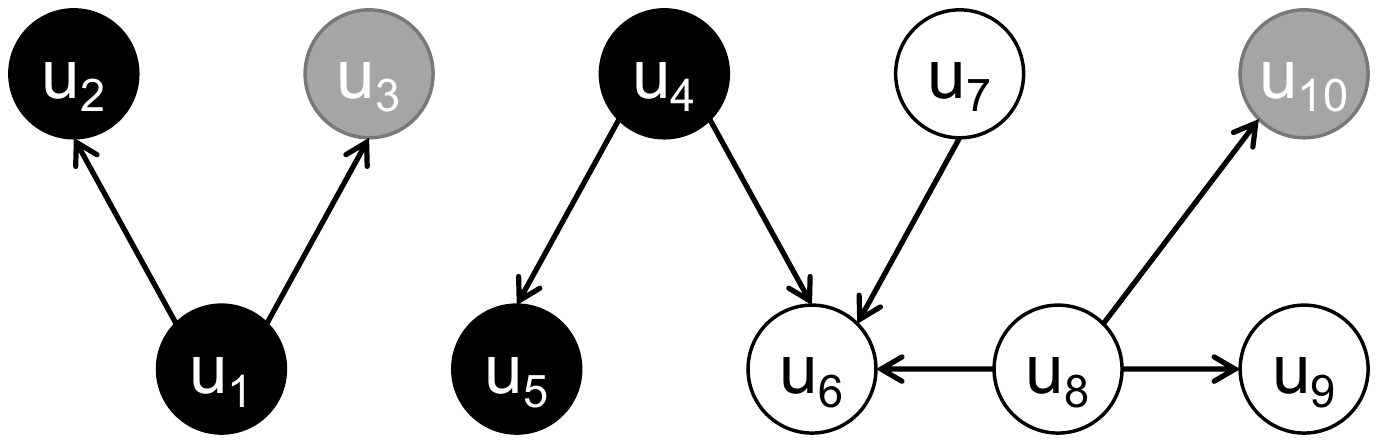}
  \caption{Counterexamples for the monotonicity and submodularity of OIM.}
  \label{fig:fig2}
\end{figure}
\begin{example}
  We present a graph $G$ with $10$ users in \cref{fig:fig2}, where the nodes in black ($u_1, u_2, u_4, u_5$), gray ($u_3, u_{10}$), and white ($u_6, u_7, u_8, u_9$) denote the sets of positive, neutral, and negative users for a target item, respectively.
  We assume that the probabilities of all edges are equal to $1$.
  For the vanilla IM problem with $k = 2$, $S_1 = \{u_1, u_8\}$ is the optimal seed set because $S_1$ activates $7$ users on $G$, which is the maximum among all size-$2$ subsets.
  But we have $\mathbb{E}[I_{G}(S_1)] = -1$ since the numbers of positive and negative users activated by $S_1$ are $2$ and $3$, respectively, and thus $S_1$ is not the optimal solution to OIM.
  Alternatively, OIM returns $S_2 = \{u_1, u_5\}$ as its optimal solution with $\mathbb{E}[I_{G}(S_2)] = 3 - 0 = 3$.
\end{example}

\noindent\textbf{Hardness of OIM:}
First, OIM is NP-hard because the vanilla influence maximization~\cite{DBLP:conf/kdd/KempeKT03} (IM) problem, which is known to be NP-hard under the IC model, is a special case of OIM when $V^{+}_{p} = V$ and $V^{0}_{p}, V^{-}_{p} = \emptyset$.
Moreover, unlike vanilla IM~\cite{DBLP:conf/kdd/KempeKT03}, the objective function $\mathbb{E}[I_{G}(\cdot)]$ of OIM is non-monotone, neither submodular nor supermodular\footnote{A set function $f: 2^V \rightarrow \mathbb{R}$ is monotone if $f(S) \leq f(T)$ for any $S \subseteq T \subseteq V$; Furthermore, a set function $f: 2^V \rightarrow \mathbb{R}$ is submodular (or supermodular) if $f(S \cup \{v\}) - f(S) \geq$ (or $\leq$) $f(T \cup \{v\}) - f(T)$ for any $S \subseteq T \subseteq V$ and $v \in V \setminus T$.} due to the existence of negative opinions.
$\mathbb{E}[I_{G}(\cdot)]$ is non-monotone because its value can decrease whenever a user who activates more negative users than positive ones in expectation is added.
Next, we can show that $\mathbb{E}[I_{G}(\cdot)]$ is neither submodular nor supermodular by providing two counterexamples from \cref{fig:fig2}:
on the one hand, since $\mathbb{E}[I_{G}(\{u_4, u_7\})] - \mathbb{E}[I_{G}(\{u_4\})] = -1 > \mathbb{E}[I_{G}(\{u_7\})] - \mathbb{E}[I_{G}(\emptyset)] = -2$, $\mathbb{E}[I_{G}(\cdot)]$ is not submodular;
on the other hand, since $\mathbb{E}[I_{G}(\{u_1, u_2\})] - \mathbb{E}[I_{G}(\{u_1\})] = 0 < \mathbb{E}[I_{G}(\{u_2\})] - \mathbb{E}[I_{G}(\emptyset)] = 1$, $\mathbb{E}[I_{G}(\cdot)]$ is also not supermodular.
Therefore, the greedy algorithm for submodular maximization problems cannot provide any theoretical guarantee for OIM anymore, which implies that OIM is more challenging than vanilla IM problems.
Thus, we propose a novel algorithmic framework for OIM in the subsequent section.

\section{Our Algorithm}
\label{sec:alg}

In this section, we describe our proposed algorithm for OIM.
We first introduce the background on the reverse reachable (RR) sets to estimate the influence spread and how they are generalized for opinion-aware influence estimation in \cref{subsec:ie}.
We then propose a seed selection algorithm for OIM in \cref{subsec:alg}.
We note that the proofs of all lemmas and theorems are omitted due to space limitations and included in a technical report [xx].

\subsection{Opinion-aware Influence Estimation}
\label{subsec:ie}

In this subsection, we consider generalizing the reverse reachable (RR) set-based methods for opinion-aware influence estimation.
Generally, since computing the influence spread $\mathbb{E}[I_{G}(\cdot)]$ exactly under the IC model is \#P-hard~\cite{DBLP:conf/kdd/ChenWW10}, we typically use Monte-Carlo simulations to obtain an unbiased estimator for the influence spread.
There have been many different simulation methods for influence estimation, among which the reverse reachable (RR) set-based method first proposed by Borgs \emph{et al.}~\cite{DBLP:conf/soda/BorgsBCL14} is the most widely used one.
After the seminal work of~\cite{DBLP:conf/soda/BorgsBCL14}, more efficient RR set-based methods~\cite{DBLP:conf/sigmod/TangXS14, DBLP:conf/sigmod/TangSX15, DBLP:conf/aaai/OhsakaAYK14, DBLP:conf/sigmod/NguyenTD16, DBLP:conf/sigmod/Guo0WC20} have been proposed for influence estimation in different IM problems.

Formally, an RR set $\mathcal{R}_u$ of a node $u \in V$ on $G$ is generated in two steps: \emph{i)} acquire the transpose graph $G^{\top} = (V, E^{\top})$ of $G$, where $(v_i, v_j) \in E^{\top}$ iff $(v_j, v_i) \in E$; \emph{ii)} simulate the diffusion process starting from $u$ on $G^{\top}$ to obtain the set of nodes activated by $u$ as the RR set $\mathcal{R}_u$.
Under the IC model, an equivalent method for RR set generation in the existing literature~\cite{DBLP:conf/aaai/OhsakaAYK14} is as follows: We generate a graph $G'$ by removing each edge $e \in E$ with probability $1 - w(e)$ independently. For any node $u$, an RR set $\mathcal{R}_u$ is the set of nodes in $G'$ that can reach $u$.
Intuitively, if a seed set $S$ has a higher probability of activating $u$, i.e., the value of $\Pr[S \leadsto u]$ is greater, there will be a higher chance that the intersection of $S$ and $\mathcal{R}_u$ is non-empty.
According to the analysis in~\cite{DBLP:conf/soda/BorgsBCL14}, we have the following lemma to formalize the above intuition.
\begin{lemma}\label{lm:rrs}
  For a seed set $S \subseteq V$, a node $u \in V$, and an RR set $\mathcal{R}_u$ of node $u$, we have $\Pr[S \leadsto u] = \Pr[\mathcal{R}_u \cap S \neq \emptyset]$.
\end{lemma}
\Cref{lm:rrs} is a special case of \cite[Observation 1]{DBLP:conf/soda/BorgsBCL14}, which implies that an unbiased estimation $\sigma_{u}(S)$ of $\Pr[S \leadsto u]$ can be obtained by estimating the probability of the event $\mathcal{R}_u \cap S \neq \emptyset$.
Next, we analyze how many RR sets are needed to get an accurate estimation $\sigma_{u}(S)$ within a small error with high confidence.
We consider $l$ instances $G'_{1}, G'_{2}, \ldots, G'_{l}$ are generated independently, each is used to compute an RR set $\mathcal{R}_{u, i}$ of node $u$ to estimate $\Pr[S \leadsto u]$ for any $S \subseteq V$.
We define a Bernoulli random variable $X_{S, i}$ for $S$ and $\mathcal{R}_{u, i}$ as:
\begin{equation}
\label{eq:ind:var}
  X_{S, i} =
  \begin{cases}
    1, \quad \text{if} \; \mathcal{R}_{u, i} \cap S \neq \emptyset \\
    0, \quad \text{otherwise}
  \end{cases}
\end{equation}
\cref{lm:rrs} implies that $\Pr[S \leadsto u] = \mathbb{E}[X_{S, i}]$.
Thus, we have the following lemma to indicate the number of RR sets needed to get an $(\varepsilon, \delta)$-approximation $\sigma_{u}(S)$ of $\Pr[S \leadsto u]$ for any $S \subseteq V$ of size at most $k$ and $u \in V$.
\begin{lemma}
\label{lm:inf:est}
  Let $l = O(\frac{k\log{n}}{\varepsilon^2} \log{\frac{1}{\delta}})$, $\mathcal{R}_{u, 1}, \ldots, \mathcal{R}_{u, l}$ be the RR sets for any $u \in V$, and $X_{S, 1}, \ldots, X_{S, l}$ be the random variables defined by \cref{eq:ind:var} for any $S \subseteq V$ with $|S| \leq k$. Then, with probability at least $1 - \delta$, it holds that $\left|\sigma_{u}(S) - \Pr[S \leadsto u] \right| < \varepsilon$, where $\sigma_{u}(S) = \frac{1}{l} \sum_{i=1}^{l} X_{S, i}$.
\end{lemma}
\begin{proof}
  First of all, we have $X_{S, i} \in [0, 1]$ for each $i \in [l]$. By applying Hoeffding's inequality~\cite{doi:10.1080/01621459.1963.10500830} on $X_{S, 1}, \ldots, X_{S, l}$, we have
  \begin{displaymath}
  	\Pr\left[ |\overline{X} - \mathbb{E}[\overline{X}]| \geq \varepsilon \right] < 2 \exp(-2l\varepsilon^2),
  \end{displaymath}
  where $\overline{X} = \sigma_{u}(S) = \frac{1}{l}\sum_{i=1}^{l} X_{S, i}$, as $\sum_{i=1}^{l}(b_i - a_i)^2 = l$.
  Furthermore, it holds that $\sigma_{u}(S) = \mathbb{E}[\overline{X}]$ from \cref{lm:rrs}.
  The probability that $\left|\sigma_{u}(S) - \Pr[S \leadsto u] \right| < \varepsilon$ is thus at least $1 - 2 \exp(-2l\varepsilon^2)$ for a specific $u$ and $S$.
  Taking the union bound over all $\sum_{i=1}^{k}\binom{n}{i} = O(n^k)$ seed sets of size at most $k$ and all $n$ nodes $u \in V$, the probability that $\left|\sigma_{u}(S) - \Pr[S \leadsto u] \right| < \varepsilon$ holds for any $S$ and $u$ is at least $1 - 2 \exp(-2l\varepsilon^2) \cdot O(n^{k+1})$.
  And we need $l = O(\frac{k\log{n}}{\varepsilon^2} \log{\frac{1}{\delta}})$ RR sets to obtain an $(\varepsilon, \delta)$-approximation $\sigma_{u}(S)$ of $\Pr[S \leadsto u]$ with probability at least $1 - \delta$.
\end{proof}

Based on \cref{lm:inf:est}, we further obtain the estimations for the expected values $\mathbb{E}[I^+_{G}(S)]$, $\mathbb{E}[I^-_{G}(S)]$ of the positive and negative influence spread functions as well as the expected value $\mathbb{E}[I_{G}(S)]$ of the opinion-aware influence function for any $S \subseteq V$ in the following lemma.
\begin{lemma}\label{lm:sigma}
  Let $ \sigma^{+}(S) := \sum_{u \in V^+_p} \sigma_{u}(S) $ and $ \sigma^{-}(S) := \sum_{u \in V^-_p} \sigma_{u}(S) $.
  We have $ |\sigma^{+}(S) - \mathbb{E}[I^+_{G}(S)]| \leq \varepsilon |V^+_p| $ and $|\sigma^{-}(S) - \mathbb{E}[I^-_{G}(S)]| \leq \varepsilon |V^-_p|$ with probabilities at least $1- \delta |V^+_p|$ and $1- \delta |V^-_p|$.
  Moreover, $ |\sigma(S) - \mathbb{E}[I_{G}(S)]| \leq \varepsilon n $ with probability at least $1- \delta n$ where $\sigma(S) = \sigma^{+}(S) - \sigma^{-}(S) $.
\end{lemma}
\begin{proof}
  The first two inequalities are obtained by summing up the $(\varepsilon, \delta)$-approx-imations for all users $ u \in V^+_p $ or $ u \in V^-_p $ in \cref{lm:inf:est} and applying the union bound on all the estimations.
  Then, we have the third inequality by taking the difference between the first two inequalities.
\end{proof}

\subsection{Seed Selection}
\label{subsec:alg}

\begin{algorithm}[tb]
  \footnotesize
  \caption{\textsc{SandwichGreedy}}
  \label{alg:ss:1}
  \begin{algorithmic}[1]
    \REQUIRE Graph $G = (V, E, w)$, user vectors $\mathbf{H}_{V} = \{ \mathbf{h}_u : u \in V \}$, target item vector $\mathbf{h}_p$, seed set size $k \in \mathbb{Z}^{+}$, parameters $\varepsilon, \delta \in (0, 1)$
    \ENSURE A set $S' \subseteq V$ such that $|S'| \leq k$
    \STATE Compute $o_{pu}$ based on \cref{eq:opinion} from $\mathbf{h}_u$ for each $u \in V$ and $\mathbf{h}_p$\label{ln:opinion}
    \STATE Let $V^{+}_{p} = \{ u \in V \,|\, o_{pu} = 1 \}$ and $V^{-}_{p} = \{ u \in V \,|\, o_{pu} = -1 \}$\label{ln:set}
    \STATE Generate a set of $l = O(\frac{k\log{n}}{\varepsilon^2} \log{\frac{1}{\delta}})$ RR sets of $G$\label{ln:rrs}
    \STATE Initialize $S, \overline{S}, \underline{S} \gets \emptyset$
    \FOR {$i = 1$ to $k$}
    \STATE Define $\Delta_{\sigma}(u \,|\, S) := \sigma(S \cup \{u\}) - \sigma(S)$, $\Delta_{\overline{\sigma}}(u \,|\, \overline{S}) := \sigma(\overline{S} \cup \{u\}) - \sigma(\overline{S})$, $\Delta_{\underline{\sigma}}(u \,|\, \underline{S}) := \sigma(\underline{S} \cup \{u\}) - \sigma(\underline{S})$ w.r.t.~$\sigma, \overline{\sigma}, \underline{\sigma}$
    \STATE Compute $\Delta_{\sigma}(u \,|\, S), \Delta_{\overline{\sigma}}(u \,|\, \overline{S}), \Delta_{\underline{\sigma}}(u \,|\, \underline{S})$ for each $u \in V$ based on the RR sets
    \STATE Find the nodes $u^*_i \gets \argmax_{u \in V} \Delta_{\sigma}(u \,|\, S)$, $\overline{u}^*_i \gets \argmax_{u \in V}$ $\Delta_{\overline{\sigma}}(u \,|\, \overline{S})$, $\underline{u}^*_i \gets \argmax_{u \in V} \Delta_{\underline{\sigma}}(u \,|\, \underline{S})$ \label{ln:greedy}
    \STATE Update the seed sets $S \gets S \cup \{u^*_i\}$, $\overline{S} \gets \overline{S} \cup \{\overline{u}^*_i\}$, $\underline{S} \gets \underline{S} \cup \{\underline{u}^*_i\}$
    \ENDFOR
    \STATE \textbf{return} $S' \gets \argmax \{\sigma(S), \sigma(\overline{S}), \sigma(\underline{S})\}$
  \end{algorithmic}
\end{algorithm}

Given the influence estimation results in \cref{subsec:ie}, we introduce our algorithm for seed selection.
First of all, for a target item $p$ with its vector representation $\mathbf{h}_p$, we can compute the positive and negative sets of users $V^{+}_{p}$ and $V^{-}_{p}$ based on $\mathbf{H}_V = \{ \mathbf{h}_u : u \in V \}$ by computing the value of $o_{pu}$ for each $u \in V$ based on \cref{eq:opinion}.
Then, we generate a set of RR sets on graph $G$ under the OIC model, from which the value of the objective function for any seed set of size at most $k$ can be estimated with bounded absolute errors as indicated in \cref{lm:sigma}.
Accompanied by the opinion-aware influence estimation based on RR sets, we should identify the seeds with high influence in $V^{+}_{p}$ and low influence in $V^{-}_{p}$ to maximize $\mathbb{E}[I_{G}(S)]$ for OIM.
The classic approach to seed selection in IM is to use the \emph{greedy strategy}, i.e., picking a node that maximally increases $\mathbb{E}[I_{G}(S)]$ progressively in $k$ iterations.
However, since the objective function $\mathbb{E}[I_{G}(S)]$, as well as its estimated function $\sigma(S)$, are non-submodular, the greedy algorithm no longer has any theoretical guarantee.
Thus, we consider extending the greedy selection procedure using the sandwich strategy to provide a data-dependent approximation for OIM.
In particular, to apply the sandwich strategy for greedy non-submodular maximization, we need to find the upper-bound and lower-bound submodular functions for $\mathbb{E}[I_{G}(S)]$ and $\sigma(S)$.
According to the analysis in~\cite{DBLP:conf/kdd/KempeKT03}, the influence spread function under the IC model is monotone and submodular, i.e., $\mathbb{E}[I^+_{G}(S)]$ and $\mathbb{E}[I^-_{G}(S)]$ are both monotone and submodular.
In addition, their estimated functions $\sigma^+(S)$ and $\sigma^-(S)$ based on RR sets can be seen as coverage functions, which are also monotone and submodular.
In this way, $\mathbb{E}[I_{G}(S)]$ and $\sigma(S)$ can be naturally written in the form of the difference of two submodular functions.
Following the difference of submodular (DS) decomposition in~\cite{DBLP:conf/uai/IyerB12}, we derive the modular (Note that a function is called a modular function if it is both submodular and supermodular) upper and lower bounds of any submodular function $f$ as follows.
Given a submodular function $f : 2^V \rightarrow \mathbb{R}$ and any subset $X \subseteq V$, the modular upper bound function $\overline{f}$ of $f$ is defined as
\begin{equation}
  \overline{f}(S) := f(X) - \sum_{u \in X \setminus S} f(u | S \setminus \{u\}) + \sum_{u \in S \setminus X} f(u | \emptyset)
\end{equation}
where $f(u | S) = f(S \cup \{u\}) - f(S)$.
Let $\pi$ be a permutation of $V$ and define $S^{\pi}_{i} = \{ \pi(1), \pi(2), \ldots, \pi(i) \}$ as $\pi$'s chain containing $X$, where $S^{\pi}_{0} = \emptyset$ and $S^{\pi}_{|X|} = X$.
Then, we define the modular lower bound function $\underline{f}$ of $f$ as
\begin{equation}
  \underline{f}(S) = \sum_{u \in S} g^{\pi}_{X}(u), \; g^{\pi}_{X}(\pi(i)) = f(S^{\pi}_{i}) - f(S^{\pi}_{i-1})
\end{equation}
Based on the above definitions, we formulate the submodular upper and lower bounds of the estimated objective function $\sigma(S)$ as $\overline{\sigma}(S) = \overline{\sigma}^{+}(S) - \underline{\sigma}^{-}(S)$ and $\underline{\sigma}(S) = \underline{\sigma}^{+}(S) - \overline{\sigma}^{-}(S)$ due to $ \underline{\sigma}^{+}(S) \leq \sigma^{+}(S) \leq \overline{\sigma}^{+}(S) $, $ \underline{\sigma}^{-}(S) \leq \sigma^{-}(S) \leq \overline{\sigma}^{-}(S) $, and the modularity of $\overline{f}$ and $\underline{f}$ for any submodular function $f$.
Having the upper and lower bounds, the sandwich strategy is to run the greedy selection procedure with respect to $\sigma, \overline{\sigma}, \underline{\sigma}$ independently in $k$ rounds to find three sets $S, \overline{S}, \underline{S}$ of seeds, respectively.
Especially, starting from empty sets $S, \overline{S}, \underline{S}$, it finds the nodes $u^*, \overline{u}^*, \underline{u}^*$ that maximize the marginal increases in $\sigma(S), \overline{\sigma}(\overline{S}), \underline{\sigma}(\underline{S})$ and adds them to $S, \overline{S}, \underline{S}$ accordingly at each round.
Finally, it returns the seed set with the largest (estimated) objective value among the three as the final result $S'$.
The above steps for seed selection are described in Algorithm~\ref{alg:ss:1}.

\smallskip\noindent\textbf{Theoretical Analysis:}
We then analyze the data-dependent approximation ratio of the seed set $S'$ provided by \cref{alg:ss:1}.
First of all, we show that \cref{alg:ss:1} finds an approximation to the best greedy selection with high probability at each iteration in the following lemma.
\begin{lemma}\label{lm:approx:greedy}
  The $i$-th node $u^*_i$ selected by \cref{alg:ss:1} satisfies that $\Delta_{\sigma}(u^*_i \,|\, S) \geq \mathbb{E}[I_{G}(S \cup \{u^*_i\}) - I_{G}(S)] - 2 \varepsilon n$ with probability at least $1- \delta n$.
\end{lemma}
\begin{proof}
  According to \cref{lm:sigma}, $ |\sigma(S) - \mathbb{E}[I_{G}(S)]| \leq \varepsilon n $ with probability at least $1 - \delta n$ for any $S \subseteq V$ of size at most $k$.
  Thus, $\sigma(S) - \mathbb{E}[I_{G}(S)] < - \varepsilon n$ and $\sigma(S) - \mathbb{E}[I_{G}(S)] > \varepsilon n$ both with probability at most $\frac{\delta n}{2}$.
  Taking $S$ and $S \cup \{u\}$ into the inequalities, we have $|\Delta_{\sigma}(u \,|\, S) - \mathbb{E}[I_{G}(S \cup \{u\}) - I_{G}(S)]| \leq 2 \varepsilon n$ with probability at least $1 - \delta n$ for any $u \in V$.
  Naturally, the above inequality holds for node $u^*_i$, and we conclude the proof.
\end{proof}

Based on the sandwich approximation strategy, we have the following data-dependent approximation factor for \cref{alg:ss:1} on OIM.
\begin{theorem}\label{thm:approx}
  Let $S'$ be the set of seeds returned by \cref{alg:ss:1}. We have $\mathbb{E}[I_{G}(S)] \geq \max\left\{ \frac{\sigma(\overline{S})}{\overline{\sigma}(\overline{S})}, \frac{\underline{\sigma}(S^*)}{\sigma(S^*)} \right\} \cdot \left( 1 - \frac{1}{e} \right) \cdot \mathbb{E}[I_{G}(S^*)] - O(\varepsilon k n)$ with probability at least $1- \delta k n$, where $S^*$ is the optimal solution to maximize $\mathbb{E}[I_{G}(\cdot)]$ under cardinality constraint $k$.
\end{theorem}
\begin{proof}
  Since $\overline{\sigma}$ and $\underline{\sigma}$ are both monotone submodular functions, we have $\overline{\sigma}(\overline{S}) \geq (1 - 1/e) \cdot \overline{\sigma}(\overline{S}^*)$ and $\underline{\sigma}(\underline{S}) \geq (1 - 1/e) \cdot \underline{\sigma}(\underline{S}^*)$, where $\overline{S}^*$ and $\underline{S}^*$ are the optimal solutions to maximize $\overline{\sigma}$ and $\underline{\sigma}$ subject to cardinality constraint $k$, respectively.
  For the solution $\overline{S}$ to the upper-bound function $\overline{\sigma}$, we have
  \begin{multline*}
    \sigma(\overline{S})
    \geq \frac{\sigma(\overline{S})}{\overline{\sigma}(\overline{S})} \cdot \left( 1 - \frac{1}{e} \right) \cdot \overline{\sigma}(\overline{S}^*) \\
    \geq \frac{\sigma(\overline{S})}{\overline{\sigma}(\overline{S})} \cdot \left( 1 - \frac{1}{e} \right) \cdot \overline{\sigma}(S^*) 
    \geq \frac{\sigma(\overline{S})}{\overline{\sigma}(\overline{S})} \cdot \left( 1 - \frac{1}{e} \right) \cdot \sigma(S^*).
  \end{multline*}
  For the solution $\underline{S}$ to the lower-bound function $\underline{\sigma}$, we have
  \begin{multline*}
    \sigma(\underline{S})
    \geq \underline{\sigma}(\underline{S})
    \geq \left( 1 - \frac{1}{e} \right) \cdot \underline{\sigma}(\underline{S}^*) \\
    \geq \left( 1 - \frac{1}{e} \right) \cdot \underline{\sigma}(S^*)
    \geq \frac{\underline{\sigma}(S^*)}{\sigma(S^*)} \cdot \left( 1 - \frac{1}{e} \right) \cdot \sigma(S^*).
  \end{multline*}
  Furthermore, by applying the result of \cref{lm:sigma} $k$ times for each node $u^*_i \in S$, we have $\sigma(S) \geq \mathbb{E}[I_{G}(S)] - 2 \varepsilon k n$ with probability at least $1- \delta k n$.
  Note that the same result also holds for $\overline{S}$, $\underline{S}$, and $S^*$.
  Thus, we conclude the proof by combining all the above results.
\end{proof}

In terms of time complexity, computing $o_{p u}$ as well as $V^{+}_{p}$ and $V^{-}_{p}$ in Lines~\ref{ln:opinion}--\ref{ln:set} takes $O(nd)$ time where $d$ is the dimensionality of $\mathbf{h}_u$ and $\mathbf{h}_p$.
Then, generating the RR sets in Line~\ref{ln:rrs} requires $O(\frac{m k\log{n}}{\varepsilon^2} \log{\frac{1}{\delta}})$ time.
Next, it runs in $k$ iterations for seed selection and takes $O(\frac{n k\log{n}}{\varepsilon^2} \log{\frac{1}{\delta}})$ time per iteration to find and add the nodes with the maximum marginal gains.
To guarantee that the approximation factor in \cref{thm:approx} satisfies with a constant probability, i.e., $\delta k n = O(1)$, we should set $1/\delta = O(n k) $.
Therefore, the overall time complexity of \cref{alg:ss:1} is $O\left( \varepsilon^{-2} (m k + n k^2) \log{n} (\log{n} + \log{k}) \right)$.

\section{Experiments}
\label{sec:exp}

To evaluate the performance of our approach, we conduct extensive experiments on real-world data sets.
Next, we introduce our experimental setup in Section~\ref{subsec:setup}.
Then, we present our experimental results in Section~\ref{subset:results}.

\subsection{Setup}
\label{subsec:setup}

\noindent\textbf{Datasets:}
We use three publicly available real-world datasets, i.e., Ciao, Epinion\footnote{\url{https://www.cse.msu.edu/~tangjili/datasetcode/truststudy.htm}}, and MovieLens\footnote{\url{https://grouplens.org/datasets}}.
Ciao and Epinion are two product review data sets containing users' ratings on products and their trust relationships.
MovieLens is a recommendation datasets consisting of user ratings on movies.
As the social relationships between users are unavailable on the MovieLens dataset, we generate an influence graph for it by adding an edge between each pair of users whose rating histories have a Jaccard similarity of greater than a threshold (e.g., $0.2$).
As a pre-processing step, we normalize the ratings of all datasets to the same scale $[0, 5]$, where higher ratings denote more positive opinions.
We summarize the statistics of the three datasets in~\cref{tab:datasets}.

\begin{table}[t]
  \scriptsize
  \centering
  \caption{Statistics of real-world datasets in the experiments.}\label{tab:datasets}
  \begin{tabular}{|l|l|l|l|l|l|}
  \hline
  Dataset & Type & \# users & \# edges & \# items & \# ratings \\
  \hline
  \hline
  Ciao & directed, product & 2,378 & 57,544 & 16,861 & 36,065 \\
  \hline
  Epinion & directed, product & 49,289 & 355,813 & 139,738 & 664,824 \\
  \hline
  MovieLens & directed, movie & 6,040 & 136,122 & 3,900 & 1,000,209 \\
  \hline
  \end{tabular}
\end{table}
\begin{figure}[t]
  \centering
  \includegraphics[width=0.4\linewidth]{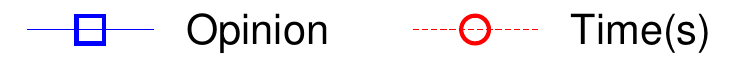}
  \vspace{-1em}
  \\
  \subfloat[Ciao \label{fig:eps:1}]{\includegraphics[width=.33\linewidth]{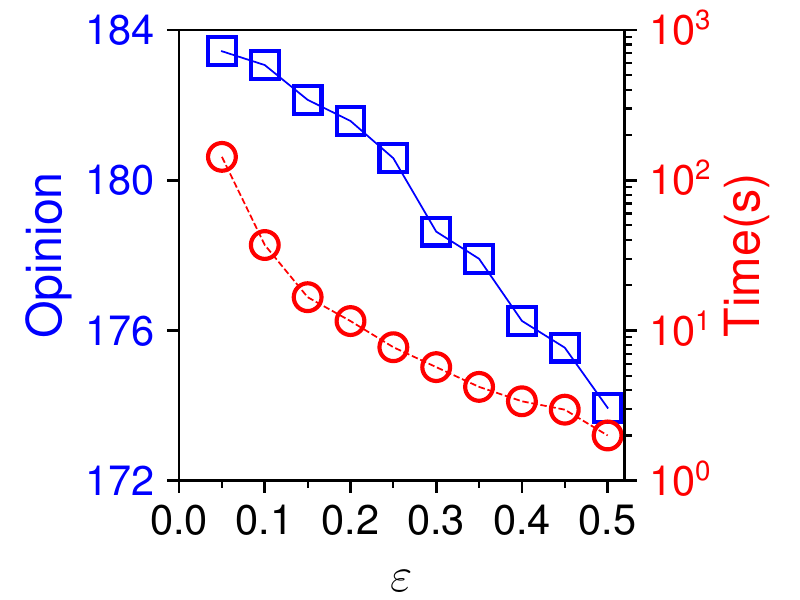}}
  \subfloat[Epinion \label{fig:eps:2}]{\includegraphics[width=.33\linewidth]{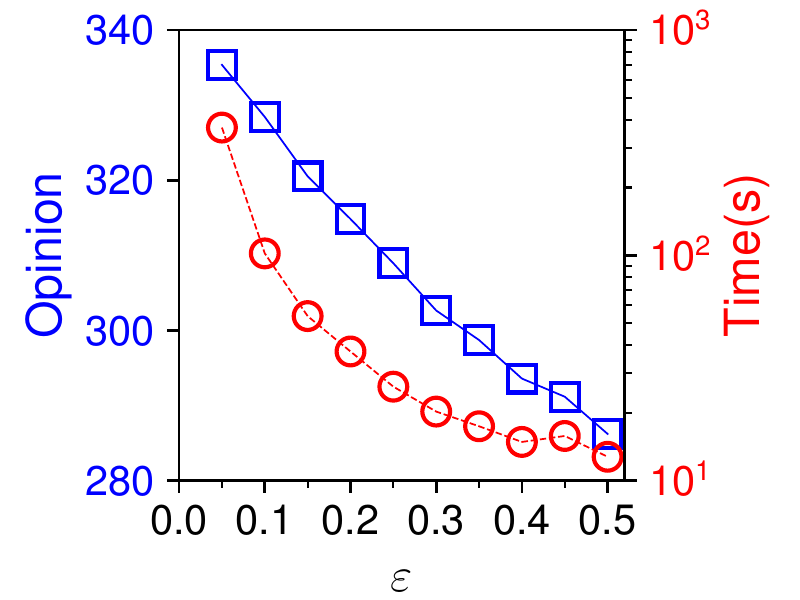}}
  \subfloat[MovieLens \label{fig:eps:3}]{\includegraphics[width=.33\linewidth]{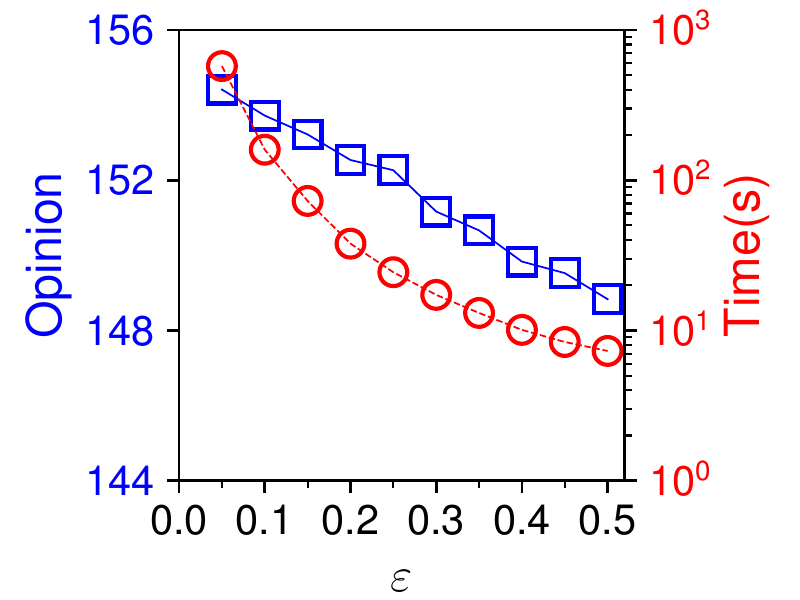}}
  \caption{Total opinions and running time of OIM by varying the error parameter $\varepsilon$.}
  \label{fig:eps}
\end{figure}
\begin{figure*}[t]
  \centering
  \includegraphics[width=0.4\linewidth]{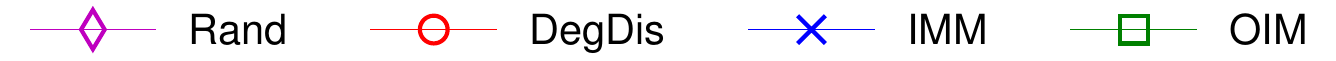}
  \vspace{-1em}
  \\
  \subfloat[Ciao \label{fig:ko:1}]{\includegraphics[width=.16\linewidth]{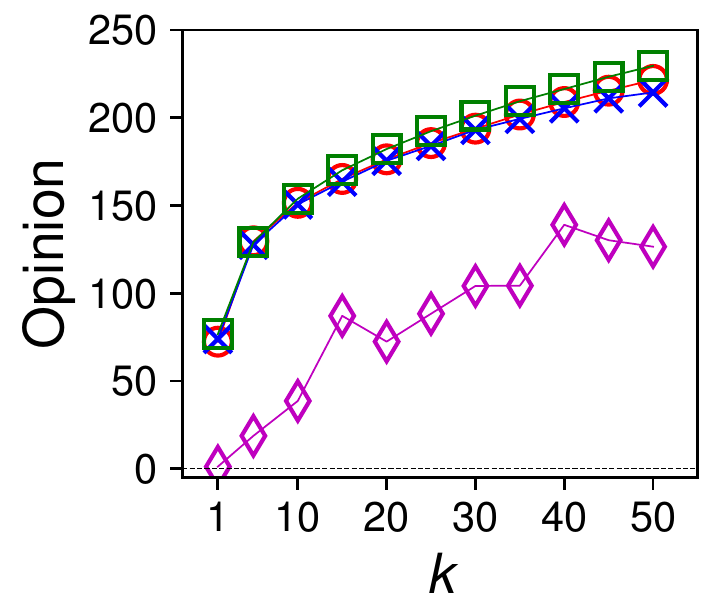}\includegraphics[width=.16\linewidth]{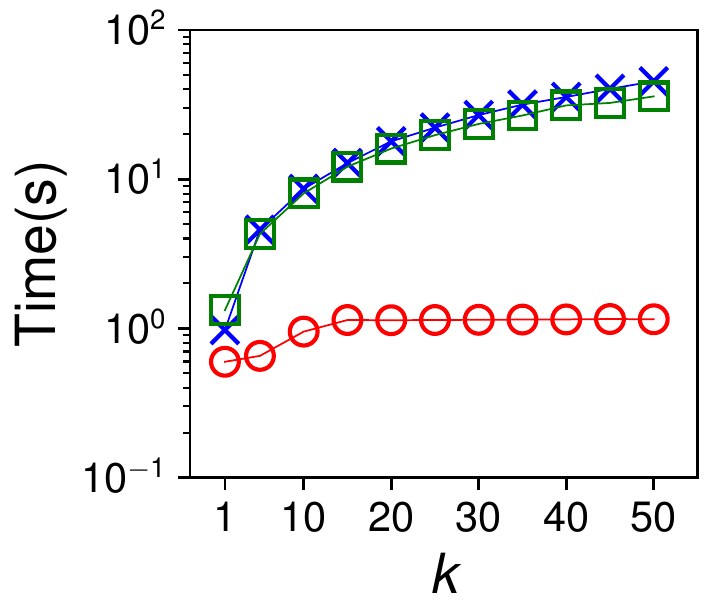}}
  \subfloat[Epinion \label{fig:ko:2}]{\includegraphics[width=.16\linewidth]{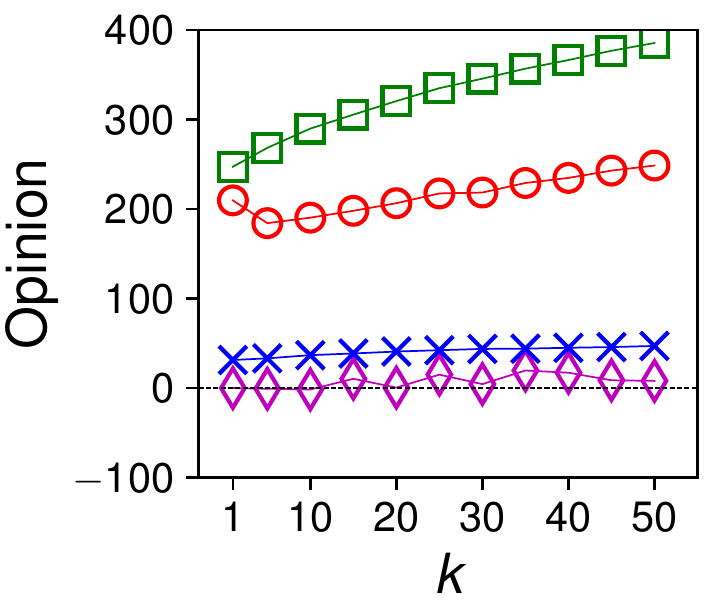}\includegraphics[width=.16\linewidth]{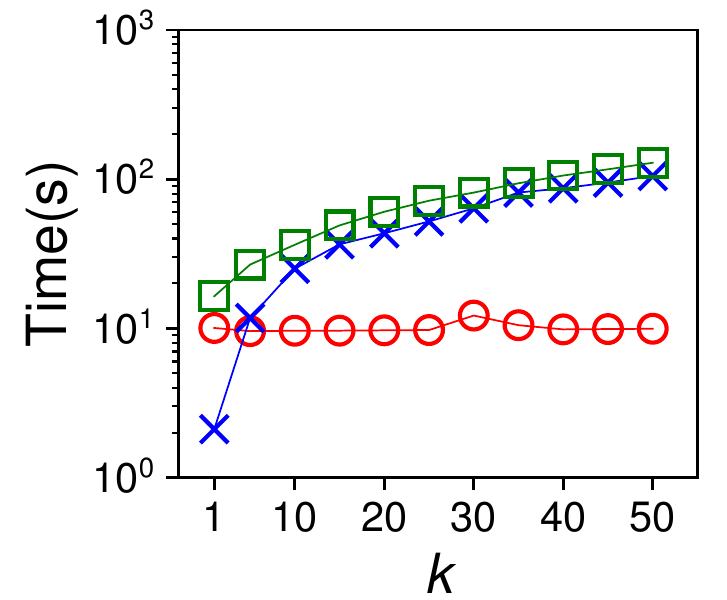}}
  \subfloat[MovieLens \label{fig:ko:3}]{\includegraphics[width=.16\linewidth]{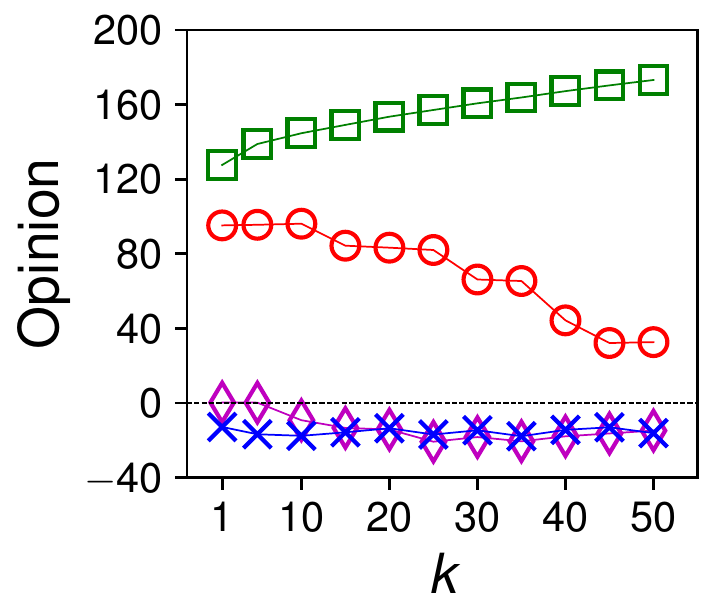}\includegraphics[width=.16\linewidth]{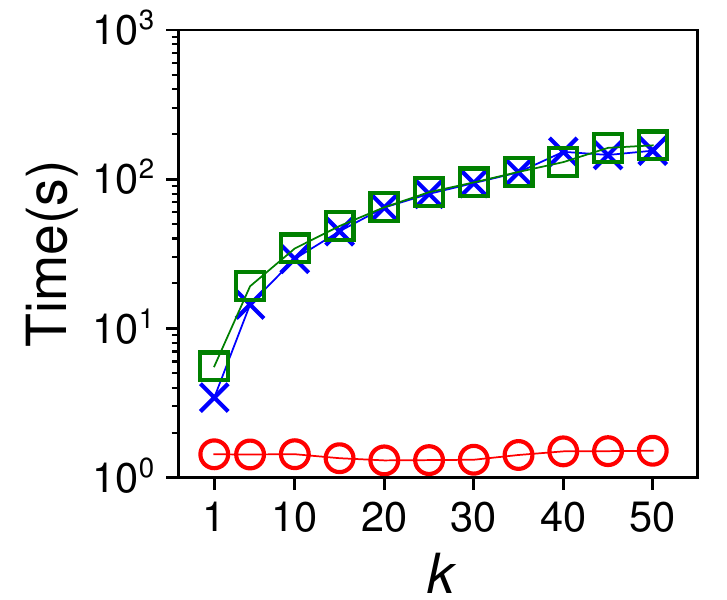}}
  \caption{Total opinions and running time of different algorithms by varying the seed set size $k$.}
  \label{fig:ko}
\end{figure*}

\smallskip\noindent\textbf{Algorithms:}
We compare our proposed OIM algorithm with the following baselines under the same OIC model used in this paper:
(1) \textsc{Rand} that randomly picks $k$ seed users; (2) degree discount (\textsc{DegDis}) heuristic~\cite{DBLP:conf/kdd/ChenWY09}; and (3) an RRS-based algorithm \textsc{IMM}~\cite{DBLP:conf/sigmod/TangSX15}.
We do not compare with the algorithms in~\cite{DBLP:conf/icdcs/ZhangDT13, DBLP:conf/kdd/LiuKY18, DBLP:conf/asunam/LuoLK19} because they are specific for LT models and cannot work under the OIC model.
Other IM algorithms are ignored since they are opinion and item-unaware, whose results are mostly close to or worse than IMM's.

\smallskip\noindent\textbf{Implementation:}
In our implementation, we extend \textsc{DegDis} to be opinion-aware by choosing the node with the largest difference between positive and negative degrees, instead of the one with the largest overall degree, in each iteration.
We use the original \textsc{IMM} algorithm as its influence estimation process cannot be generalized to the OIC model.
On each dataset, we utilize an existing GNN-based model~\cite{DBLP:conf/www/Fan0LHZTY19} to embed all users and items into the same vector space to estimate the (unobserved) ratings of users on items.
Then, the weight $w(e)$ of each edge $e = (u, v)$ is computed based on the cosine similarity between their vector representations $\mathbf{h}_{u}$ and $\mathbf{h}_{v}$.
Moreover, we use the average of all observed ratings in each dataset as the neutral rating value $r_{0}$ and set the parameter $\tau$ to $0.5$.
Finally, for the seed set $S$ computed by each algorithm, we run 1,000 Monte Carlo simulations to estimate its total opinion, i.e., $\mathbb{E}[I_{G}(S)]$.

All algorithms were implemented in Python 3, and PyTorch was used for GNN training and inference.
The experiments were run on a server with an Intel Xeon E5-2650v4 2.2GHz processor, 96GB main memory, and an NVIDIA Tesla V100 GPU with 16GB HBM2 memory, running Ubuntu 18.04 LTS.


\subsection{Experimental Results}
\label{subset:results}

\noindent\textbf{Effect of Error Parameter $\varepsilon$:}
We present the total opinion (i.e., difference between the numbers of positive and negative users activated by the seed set) and running time of our OIM algorithm on each dataset by varying the parameter $\varepsilon$ from $0.05$ to $0.5$ in~\cref{fig:eps}.
Here, we follow existing RR set-based methods~\cite{DBLP:conf/sigmod/TangSX15,DBLP:conf/sigmod/TangXS14} to set $\delta = 100kn$.
The size $k$ of the seed set is fixed to $10$.
Generally, we observe that the total opinion and running time of OIM decrease with an increasing $\varepsilon$ across all datasets.
Based on our analysis in \cref{sec:alg}, the number of RR sets for OIM is linear with respect to $\varepsilon^{-2}$.
Accordingly, on the one hand, a larger $\varepsilon$ leads to a less stable influence estimation, a greater error in seed selection, and thus slightly lower seed quality (cf.~Lemmas~\ref{lm:inf:est}--\ref{lm:approx:greedy}); on the other hand, a quadratically fewer number of RR sets w.r.t.~$\varepsilon$ also significantly improves the time efficiency of OIM.
In the remaining experiments, we use $\varepsilon = 0.15$ and $\delta = 100kn$ to determine the number of RR sets to sample in the OIM algorithm to strike a good balance between seed quality and time efficiency.

\smallskip\noindent\textbf{Effect of Solution Size $k$:}
We show the total opinion and running time of each algorithm by varying the size $k$ of the seed set from $1$ to $50$ in Fig.~\ref{fig:ko}.
We randomly sample 50 items from each dataset, select a size-$k$ seed set for each item using different algorithms, and report the average opinion and running time of each algorithm across all items.
Our OIM algorithm always provides seed sets with the most positive opinions among all algorithms in all cases.
Nevertheless, its advantage over other algorithms depends on the item's opinion distribution. On the Ciao dataset, since almost all items receive more positive opinions than negative ones, OIM is only marginally better than \textsc{DegDes} and IMM.
On all other three datasets, as positive and negative opinions are generally balanced, the total opinions of \textsc{Rand} and IMM are close to or lower than $0$, as they are opinion-unaware. However, OIM still provides seed sets with more positive opinions, thus achieving significant improvements upon all baselines in terms of seed quality.
In terms of time efficiency, the running time of OIM is generally close to that of IMM, which both much slower than that of \textsc{DegDes} and \textsc{Rand} for larger $k$'s.
This is because the numbers of RR sets in OIM and IMM are close to each other when the same value of $\varepsilon$ is used, and the time for seed selection is nearly negligible compared to the time for RR set generation for both OIM and IMM.
Accordingly, \textsc{DegDes} and \textsc{Rand} run much faster than OIM and IMM because they do not require the time-consuming RR set generation process, yet come at the expense of lower seed quality.

\section{Conclusion}
\label{sec:conclusion}

In this paper, we defined the opinion-aware influence maximization (OIM) problem to find a set of $k$ seeds from an online social network to maximize the difference between the positive and negative opinions for a target item.
Despite its NP-hardness and non-submodularity, we proposed an efficient data-dependent approximation algorithm for OIM based on the RR sets for opinion-aware influence estimation and the sandwich approximation for seed selection.
Experimental results on three real-world datasets showed the effectiveness and efficiency of our proposed algorithm.
In future work, we would like to extend the OIM problem to more general settings where the diffusion model and user opinions can be unknown in advance and should be learned from observations.

\bibliographystyle{IEEEtran}
\bibliography{ref}

\end{document}